\documentclass[aps,prl,reprint,groupedaddress,showpacs]{revtex4-1}
\usepackage{graphicx}
\usepackage{dcolumn}
\usepackage{bm}
\usepackage[latin1]{inputenc}

\begin{document}

\title{Transverse electric surface mode in atomically thin Boron-Nitride}

\author{Michele Merano}
\email[]{michele.merano@unipd.it}

\affiliation{Dipartimento di Fisica e Astronomia G. Galilei, Universit$\grave{a}$ degli studi di Padova, via Marzolo 8, 35131 Padova, Italy}

\date{\today}
 
\begin{abstract}
The spatial confinement and the propagation length of surface waves in a single-layer two-dimensional atomic crystal are analysed in term of its surface susceptibility and its surface conductivity. Based on the values of these macroscopic parameters, extracted from experimental observations, it is confirmed that graphene supports a transverse magnetic non-radiating surface mode in the ultraviolet spectral region while a single-layer hexagonal Boron-Nitride is predicted to support a transverse electric non-radiating surface mode in the visible spectrum. This last mode, at a vacuum wavelength of 633 nm, has a spatial confinement of 15 microns and an intensity-propagation distance greater than 2 cm.    
\end{abstract}


\maketitle
Surface electromagnetic waves have attracted lot of attention because of their fundamental  interest and their possible technological impact \cite{Atwater07}. The transverse evanescence of these modes renders them suitable for a broad range of applications because of the possibility to guide them along an interface. Recently the research in surface electromagnetic modes has been further boosted by the advent of two-dimensional (2D) atomic crystals \cite{Novoselov2005}. These materials are essentially single or few atomic planes pulled out of a bulk crystal. They are stable under ambient conditions, exhibit high crystal quality, and are continuous on a macroscopic scale \cite{Novoselov2005}.

In particular single-layer 2D atomic crystals have astonishing optical properties \cite{Nair2008, Heinz2010}. Their linear and non-linear optical response shows that they behave as zero-thickness interfaces \cite{Merano16, Merano15, Merano216}. In analogy to a bulk material for which it is possible to define an electrical susceptibility and a conductivity, it is possible to characterize a single-layer 2D crystal in terms of its surface electrical susceptibility and its surface conductivity. Exactly as for a 3D material these two quantities are experimentally accessible via ellipsometry \cite{Merano16, Kravets2010}. Optical absorption or optical contrast measurements \cite{Merano16, Nair2008, Blake2007} are other experimental techniques that can be used to partially fix them.  

Two-dimensional crystals can support surface electromagnetic waves. It has been predicted \cite{Vafek06, Wunsch06, Hanson08} and experimentally shown that highly confined surface plasmons can propagate on graphene \cite{Eberlein08, Koppens12, Basov12}. Apart from a localized transverse magnetic (TM) mode it was predicted that graphene can support also a transverse electromagnetic (TE) mode \cite{Ziegler07}. To date the existence of such a mode has only been experimentally proved in multi-layer graphene but not on a single-layer atomic crystal \cite{Park16}.     

Owing to the 2D nature of the collective excitations it was confirmed that the confinement of surface plasmons in graphene is much stronger than that of metallic surface plasmons. This means that graphene is ideally suited to confine light down to extremely small volumes. Also because of a large wave vector mismatch of graphene plasmons compared to free-space light, plasmon excitation and detection by light is very inefficient. In spite of that marvelous experiments have successfully achieved it \cite{Koppens12, Basov12}. Real space imaging of propagating graphene plasmons has permitted the measurement of their propagation length. One of the most appealing advantages of graphene plasmonics is the possibility to electrically control the confinement and the propagation length of surface plasmons. Experiments \cite{Koppens12, Basov12} have indeed confirmed the theoretical predictions that these quantities depend on doping \cite{Jablan09, Wang08, Polini08}.

Thus far strong surface plasmon damping has been observed on a graphene interface. Damping can be reduced when graphene-hexagonal-Boron-Nitride heterostructures \cite{Geim14} are considered. Propagation lengths of hundreds of nanometers have been observed for a vacuum exciting wavelength of 10 microns \cite{Koppens15}. Even if this is still a strong sub-wavelength damping, plasmonics in these heterostructures preserves the high field confinement typical of the 2D systems.


Infrared nano-imaging has also been used to study volume confined phonon-polaritons in a flat slab of hexagonal Boron-Nitride (BN) \cite{Hillenbrand15, Taubner15}. The measured dispersion of polaritonic waves was shown to be governed by the crystal thickness according to a scaling law that persists down to a few atomic layers \cite{Basov14, Wang15}. A surface phonon-polariton propagating within a three layer thin flake of hexagonal BN was reported at an angular frequency around 1550 cm-1 \cite{Basov14}. Even in this case anyway a strong damping was observed.     

Here I treat the surface electromagnetic modes of a single-layer 2D atomic crystal in terms of its surface conductivity and its surface electrical susceptibility. I show how the radiative or non-radiative character of these modes can be easily deduced from these quantities. Two specific examples: conducting graphene and insulating single-layer hexagonal BN will be discussed, based on the values of these macroscopic measurable parameters extracted from experimental observations. Boron-Nitride is predicted to support a TE mode with a very long propagation distance in the visible spectrum.      

I follow an approach similar to the one used by Raether \cite{Raether} for treating surface plasmon polaritons on a metallic surface. Consider a flat single-layer 2D crystal located at the interface in between two dielectric media (Fig.1), on which a TE or a TM surface electromagnetic wave propagates in the $y$ direction. In the two half-spaces ($1$) and ($2$) separated by the crystal the electric field for the TE mode is given by:
\begin{eqnarray}
\label{TE}
&\vec{\textbf{\emph{E}}}_{1}&(x, y, z, t)=(E_{x1}, 0, 0)e^{i(\omega t-k_{y1}y-k_{z1}z)} \qquad z<0 \quad \\
 \nonumber 
&\vec{\textbf{\emph{E}}}_{2}&(x, y, z, t)=(E_{x2}, 0, 0)e^{i(\omega t-k_{y2}y-k_{z2}z)} \qquad z>0  \nonumber 
\end{eqnarray}      
and the magnetic field for the TM mode by:
\begin{eqnarray}
\label{TM}
&\vec{\textbf{\emph{H}}}_{1}&(x, y, z, t)=(H_{x1}, 0, 0)e^{i(\omega t-k_{y1}y-k_{z1}z)} \qquad z<0 \quad \\
 \nonumber 
&\vec{\textbf{\emph{H}}}_{2}&(x, y, z, t)=(H_{x2}, 0, 0)e^{i(\omega t-k_{y2}y-k_{z2}z)} \qquad z>0  \nonumber 
\end{eqnarray}

\begin{figure}
\includegraphics{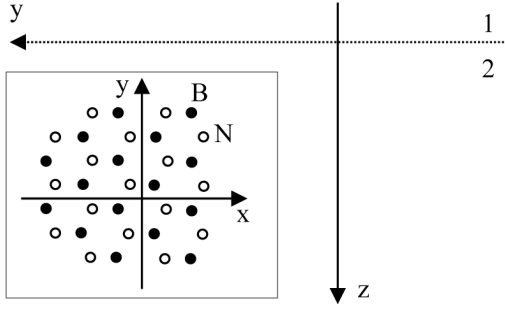}
\caption{\label{} A single-layer atomic crystal (hexagonal BN for instance) is located at the interface in between two dielectric media 1 and 2. Inset: the crystal plane}
\end{figure}

where $\omega$ is the angular frequency of the light and $k_{y}$ and $k_{z}$ are the components of the wave vector for the surface modes. These fields have to fulfill Maxwell's equations and in particular the magnetic field for the TE mode is given by:
\begin{eqnarray}
\label{Maxwell1}
rot\vec{\textbf{\emph{E}}}_{m}=-\mu_{0}\frac{\partial \vec{\textbf{\emph{H}}}_{m}}{\partial t}
\end{eqnarray}  
and the electric field for the TM mode is given by:
\begin{eqnarray}
\label{Maxwell2}
rot\vec{\textbf{\emph{H}}}_{m}=\epsilon_{0}\epsilon_{m}\frac{\partial \vec{\textbf{\emph{E}}}_{m}}{\partial t}
\end{eqnarray}
where $m$ can be $1$ or $2$. 
The boundary condition for the electric field is:
\begin{eqnarray}
\label{Boundary1}
\hat{\kappa} \wedge (\vec{\textbf{E}}_{2}-\vec{\textbf{E}}_{1})=0
\end{eqnarray}
and for the magnetic field:
\begin{eqnarray}
\label{Boundary2}
\hat{\kappa} \wedge (\vec{\textbf{H}}_{2}-\vec{\textbf{H}}_{1})=\vec{\textbf{J}}
\end{eqnarray}
where $\hat{\kappa}$ is the unit vector along the $z$ axis and $\vec{\textbf{J}}$ is the sum of the surface polarization current $\vec{\textbf{J}}_{P}$ plus the surface conduction current $\vec{\textbf{J}}_{\sigma}$ in the crystal plane: 
\begin{eqnarray}
\label{Current}
\vec{\textbf{J}}=\vec{\textbf{J}}_{P}+\vec{\textbf{J}}_{\sigma}=(i\epsilon_{0} \chi \omega+\sigma)((\hat{\kappa}\wedge \vec{\textbf{E}}_{m})\wedge \hat{\kappa})
\end{eqnarray}
where, because of (\ref{Boundary1}), $\vec{\textbf{J}}$ does not depend on $m$ and where $\epsilon_{0}$ is the vacuum permittivity, $\chi$ the electric surface susceptibility of the 2D crystal, and $\sigma$ its surface conductivity \cite{Merano16}. Finally the relation:
\begin{eqnarray}
\label{magnitude}
k^2_{ym}+k^2_{zm}=\epsilon_{m}k^2
\end{eqnarray} 
must be considered, where $k$ is the magnitude of the wave vector of light in vacuum. From (\ref{TE}), (\ref{Maxwell1}), (\ref{Boundary1}), (\ref{Boundary2}), (\ref{Current}) and (\ref{magnitude}) I obtain for the TE mode:
\begin{eqnarray}
\label{TEkz}
&k_{z1}&=\frac{k}{2}\left(\frac{\epsilon_{1}-\epsilon_{2}}{ik \chi+\sigma \eta}+(ik \chi+\sigma \eta)\right) \qquad (TE) \\
&k_{z2}&=\frac{k}{2}\left(\frac{\epsilon_{1}-\epsilon_{2}}{ik \chi+\sigma \eta}-(ik \chi+\sigma \eta)\right) \nonumber \\
&k_{y1}&=k_{y2}=\pm \sqrt{\epsilon_{1}k^{2}-k^{2}_{z1}} \nonumber
\end{eqnarray}
where $\eta$ is the impedance of vacuum and the $\pm$ symbol indicates that two counter-propagating directions are possible. From (\ref{TM}), (\ref{Maxwell2}), (\ref{Boundary1}), (\ref{Boundary2}), (\ref{Current}) and (\ref{magnitude}) I obtain for the TM mode that $k_{z2}$ is a solution of the following quartic equation:
\begin{widetext}
\begin{equation}
\label{TMkz1}
a^2k^4_{z2}+2a\epsilon_{2}kk^3_{z2}+k^2(\epsilon_{2}-\epsilon_{1}) \Big( (\epsilon_{2}+\epsilon_{1}-a^2)k^2_{z2}-2a\epsilon_{2}kk_{z2}-k^2\epsilon^2_{2} \Big) = 0 \qquad (TM)
\end{equation}
\end{widetext}
where $a=ik\chi + \sigma \eta$, and $k_{z1}$ and $k_{ym}$ are given by: 
\begin{eqnarray}
\label{TMkz2}
&k_{z1}&=\frac{k\epsilon_{1}k_{z2}}{k\epsilon_{2}+(ik\chi + \sigma \eta)k_{z2}} \qquad (TM) \\
&k_{y1}&=k_{y2}=\pm \sqrt{\epsilon_{1}k^{2}-k^{2}_{z1}}  \nonumber
\end{eqnarray} 

It is instructive to consider the simple case $\epsilon_{1}=\epsilon_{2}=1$. For the TE mode I obtain: 
\begin{eqnarray}
\label{TEkzepsilonequal}
&k_{z1}&=-k_{z2}=\frac{k}{2}\left(ik\chi + \sigma \eta\right) \qquad (TE) \\
\label{RekyTE}
&Re& (k_{y}) \approx \pm k\sqrt{1+\frac{k^2\chi^2-\sigma^2 \eta^2}{4}}  \\
\label{ImkyTE}
&Im& (k_{y}) \approx \mp k \frac{k\chi \sigma \eta}{4} 
\end{eqnarray} 
and for the TM mode:
\begin{eqnarray}
\label{TMkzepsilonequal}
&k_{z1}&=-k_{z2}=2k\frac{-ik\chi + \sigma \eta}{k^2\chi^2+\sigma^2 \eta^2} \qquad (TM) \\
\label{RekyTM}
&Re& (k_{y}) \approx \pm k\sqrt{1+\frac{4k^2\chi^2}{(k^2\chi^2+\sigma^2 \eta^2)^2}}  \\
\label{ImkyTM}
&Im& (k_{y}) \approx \pm sgn(\chi) k \frac{2\sigma \eta}{k^2\chi^2+\sigma^2 \eta^2} 
\end{eqnarray}
where $Re$ and $Im$ are the real and the imaginary part of a complex number, $sgn$ is the sign function that extracts the sign of its argument, and where the approximation sign is valid under the assumptions $k\chi<<1$ and $\sigma \eta <<1$.

In agreement with ref. \cite{Hanson08}, formula (\ref{TEkzepsilonequal}) confirms that if $\chi<0$ the TE surface mode is not a proper solution because it is exponentially growing in the $z$ direction. If $\chi>0$,  from formulas (\ref{TEkzepsilonequal}) and (\ref{RekyTE}) a non-radiating, spatially confined TE surface mode exists if $k^2\chi^2 -\sigma^2 \eta^2 >0$. This has not been noted in \cite{Hanson08}. In totally agreement with \cite{Hanson08} formulas (\ref{TMkzepsilonequal}) and (\ref{RekyTM}) confirm that a non-radiating, spatially confined TM mode is possible only for $\chi<0$.  

Based on experimental values of $\chi$ and $\sigma$ I show now that graphene supports a proper TM surface wave mode in the ultraviolet spectrum and a single-layer hexagonal BN supports a proper TE surface wave mode in the visible spectrum. The spatial confinement and the propagation distance of these modes are reported. The $\chi$ and the $\sigma$ for graphene in the visible and ultraviolet spectral range have been determined in \cite{Merano16}. The $\chi$ and the $\sigma$ for a single-layer hexagonal BN will be extracted here from published experimental data \cite{Blake2011}. 


For graphene in the spectral range 450 nm $<\lambda<$ 750 nm we have $\chi=8\cdot10^{-10}$ m and $\sigma=6\cdot10^{-5}\ \Omega^{-1}$. If we consider a free-standing graphene film, from these values and formula (\ref{RekyTE}) it is clear that the TE mode in the visible part of the spectrum is a radiative one ($Re(k_{y})<k$). The TM mode (formula (\ref{TMkzepsilonequal})) is not spatially confined in the $z$ direction. In the same paper values of $\chi=-1.2 \cdot 10^{-9}$ m and of $\sigma=18.6\cdot 10^{-5} \ \Omega^{-1}$ at a wavelength of 270 nm are reported. In this second case the TE mode is not spatially confined in the $z$ direction. The TM mode is a proper one with a spatial confinement of the field \cite{Raether} $1/\left|Im(k_{z})\right|= 4.4$ nm and an intensity propagation length \cite{Raether} of $1/\left| 2\cdot Im(k_{y})\right| =8.7\cdot 10^{-1}$ nm. This mode has already been observed in ref. \cite{Eberlein08}.

I treat now the case of a single-layer hexagonal BN. From ref. \cite{Blake2011} that reports optical contrast measurements of BN on top of a $\rm SiO_{2}/Si$ wafer with a $\rm SiO_{2}$ thickness of 290 nm, and the analysis used in ref. \cite{Merano16} it is possible to extract the value of $\chi = (1.3 \pm 0.1) \cdot 10^{-9} $ m and an upper limit for $\sigma \leq 2\cdot 10^{-6} \ \Omega^{-1}$. Figure 2 reports the experimental data published in Fig (2) of ref. \cite{Blake2011} that have been extracted from the original paper via software digitization. The observed contrast is a non-monotonic function of $\lambda$ and changes its sign at 530 nm. From simulations it emerges that the optical contrast for BN is sensitive to the sign of $\chi$. If the sign of $\chi$ is reversed the sign of the contrast is reversed. The maximum and the minimum values of the contrast increase with the magnitude of $\chi$. Increasing $\sigma$ shifts the curve upwards. This is the reason why for graphene the contrast is either positive or negligible \cite{Blake2007}. The spectral position for which the contrast changes sign depends a lot on the substrate and not on $\chi$ or $\sigma$. The two fits reported in fig. 2 are for the same best values of $\chi$ and $\sigma$. One is for the nominal thickness of 290 nm $\rm SiO_{2}$ reported in \cite{Blake2011} and the best one for a thickness of 270 nm $\rm SiO_{2}$. No other experimental data have been found in literature to better fix $\chi$ and $\sigma$. 

\begin{figure}
\includegraphics{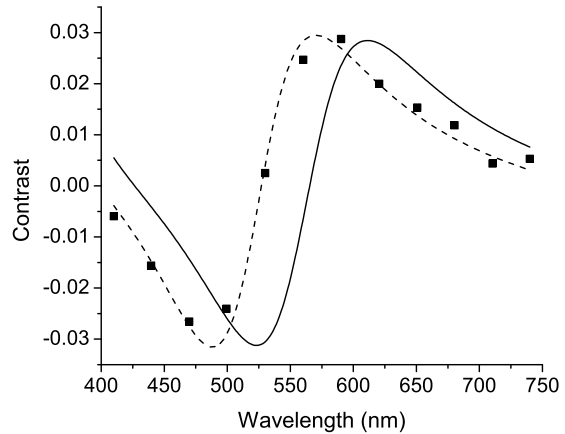}
\caption{\label{} Optical contrast as a function of the incident wavelength for a single-layer BN on top of a $\rm SiO_{2} / Si$ wafer (290 nm $\rm SiO_{2}$). Dots are experimental data extracted from ref. \cite{Blake2011}. The solid line is the best fit assuming $\chi = 1.3\cdot 10^{-9}$ m, $\sigma \leq 2\cdot 10^{-6}\ \Omega^{-1}$ and a $\rm  SiO_{2}$ thickness of 290 nm. Varying $\chi$ or $\sigma$ does not improve in any way the fit. The only way to improve the fit is by varying the $\rm SiO_{2}$ thickness, showing that the spectral position of the optical contrast curve depends much on the substrate. Dash line is the fit for the same values of $\chi$ and $\sigma$ but a $\rm SiO_{2}$ thickness of 270 nm.}

\end{figure}

Even if I choose the upper possible value for $\sigma = 2\cdot 10^{-6} \ \Omega^{-1}$, from formulas (\ref{TEkzepsilonequal}) and (\ref{RekyTE}) a free-standing, single-layer BN supports a  non-radiative TE surface mode in the visible part of the spectrum. The spatial confinement of the field for a wavelength of 633 nm is 15 $\mu$m, (a weakly localized mode) and the intensity propagation length is 2 cm, a surprising macroscopic distance. For a smaller conductivity, that it is not possible to exclude here, this propagation distance will be proportionally longer (formula \ref{ImkyTE}). This mode exists not only for a free-standing crystal but also if a single-layer BN is embedded in a dielectric (formulas \ref{TEkz}). The TM surface mode is not spatially confined in the $z$ direction.


The validity of the equations here reported is not limited to graphene or single-layer BN. They can be applied to any single-layer 2D crystal, for instance transition-metal dichalcogenides. By elucidating the role of $\chi$ and $\sigma$ in the properties of surface electromagnetic waves in 2D materials, this paper may profit to the experimental research in this field.

Based on experimental values of $\chi$ and $\sigma$, I have shown that graphene supports a non radiating TM surface mode in the ultraviolet spectral region while a single-layer hexagonal BN supports  a non-radiative TE surface mode in the visible spectrum. The two modes have very different properties. The TM mode has a very high spatial confinement and an exceedingly short propagation distance. The TE mode has a weak spatial confinement and a long propagation distance. 

While TM surface waves in graphene have been observed both in the ultraviolet \cite{Eberlein08} and in the infrared electromagnetic spectrum \cite{Koppens12, Basov12}, a TE surface mode has never been observed on any single-layer two-dimensional atomic crystal \cite{Mason14, Sipe15}. Boron-Nitride can be a good candidate for its observation in the visible spectrum. When compared with the TM mode in graphene, this TE mode has a giant propagation length. This makes it interesting both from a fundamental point of view as well as for technical applications.

%
%

\bibliography{letter}
\end{document}